\documentclass[aps,preprint,amsmath,amssymb]{revtex4-1}
\usepackage{graphicx}% Include figure files
\usepackage{here}
\begin{document}
\title{Axis thermal expansion switching in transition-metal zirconides $Tr$Zr$_2$ by tuning the $c/a$ ratio}
\author{Hiroto Arima$^{1}$}
\author{Md. Riad Kasem$^{1}$}
\author{Yoshikazu Mizuguchi$^{1}$\footnote{E-mail : mizugu@tmu.ac.jp}}
\affiliation{$^{1}$ Department of Physics, Tokyo Metropolitan University, Hachioji, Tokyo, 192-0397, Japan}
\begin{abstract}
This study examines the temperature-dependent evolution of the lattice constants for various CuAl$_2$-type compounds, including NiZr$_2$, (Co,Rh,Ir)Zr$_2$, (Fe,Co,Rh,Ir)Zr$_2$, and (Co,Ni,Cu,Rh,Ir)Zr$_2$, in the pursuit of negative or zero thermal expansion. Results reveal that NiZr$_2$ has positive thermal expansion, while the other compounds exhibit uniaxial negative thermal expansion along the $c$-axis contraction. The study suggests that the $c$-axis thermal expansion can be controlled by manipulating the $c/a$ ratio through $Tr$-site substitution, providing a design principle for achieving negative thermal expansion of the $c$-axis and potentially zero thermal expansion in a single compound in $Tr$Zr$_2$ compounds.
\end{abstract}
\maketitle
Materials possessing zero thermal expansion (ZTE) are important for certain applications, however, the attainment of ZTE materials necessitates the utilization of ultra-precise positive thermal expansion (PTE) and negative thermal expansion (NTE) materials. It is well-known that heat causes thermal expansion in almost all compounds, whereas some exhibit shrinkage upon heating. NTE compounds have been developed based on phase transitions, (such as structural\cite{Evans_2000,Amos_2001}, ferroelectric\cite{Agrawal_1988,Agrawal_1987}, magnetic transitions\cite{Schlosser_1971, Nikitin_1991}), intermetallic charge transfers\cite{Yamada_2013,Nishikubo_2019}, and valence crossovers\cite{Qiao_2021,Azuma_2007}. The significant NTE exhibited by these compounds is often limited to specific temperatures, such as those at the boundary of phase transition or crossover temperature, whereas PTE is observed over a broader range of temperatures. ZrW$_2$O$_8$ has been extensively studied for its isotropic NTE over a broad temperature range, from 0.3 K to 1000 K\cite{Martinek_1968, Mary_1996}. This isotropic NTE is disscussed to be caused by the vibration of two rigid units, a ZrO$_6$ octahedron and WO$_4$ tetrahedron, referred to as the rigid unit mode\cite{Giddy_1993, Pryde_1996, Ernst_1998, Dove_1997}.\par
Recently, transition-metal zirconides CoZr$_2$ and alloyed $Tr$Zr$_2$ (where $Tr$ denotes Fe, Co, Ni, Rh and Ir) have been reported as a new family of NTE compounds\cite{Mizugu_2022_1,Kasem_2022}. The crystal structure of $Tr$Zr$_2$ is depicted in Figure 1(a), which is a body-centered tetragonal structure of CuAl$_2$ type, in space group \#140. Neutron powder diffraction (NPD) and X-ray diffraction (XRD) at low and high temperatures reveal an anisotropic linear thermal expansion of the $a$- and $c$-axes in CoZr$_2$. In more details, it was observed that heating induces the expansion of the lattice constant $a$ (PTE) and contraction of the lattice constant $c$ (NTE), across a wide temperature range of 7 K to 573 K. Similar wide temperature range uniaxial linear thermal expansion has been reported for (Fe,Co,Ni)Zr$_2$, (Fe,Co,Ni,Rh,Ir)Zr$_2$, and (Fe,Co,Ni,Cu,Ga)Zr$_2$, which were synthesized based on the concept of high-entropy alloys (HEA). To investigate the temperature-dependent lattice parameters of solid solutions, we selected $Tr$Zr$_2$ compounds with space group \#140 ($Tr$=Fe, Co, Ni, Rh, and Ir) that have been previously reported\cite{Kuzuma_1966, Kirkpatrick_1961, Havinga_1972, Eremenko_1980}. Additionally, we included $Tr$=Cu, which possesses $d$ electrons as these elements. Our experimental analysis involved synthesizing uniaxial NiZr$_2$, (Co,Rh,Ir)Zr$_2$, (Fe,Co,Rh,Ir)Zr$_2$, and (Co,Ni,Cu,Rh,Ir)Zr$_2$ compounds and conducting XRD experiments to explore the temperature dependence of the lattice parameters for the identification of NTE compounds.\par
The polycrystalline samples of NiZr$_2$, (Co,Rh,Ir)Zr$_2$, (Fe,Co,Rh,Ir)Zr$_2$ and (Co,Ni,Cu,Rh,Ir)Zr$_2$ were synthesized by the arc melting method in an Ar gas atmosphere, on a water-cooled copper hearth. High purity powders of Ni (99.9 \%), Co (99 \%), Rh (99.9 \%), Fe (99.99 \%), Cu (99.9 \%), and Ir (99.9 \%) were pelletized and melted with Zr plates (99.2 \%). During arc melting, the ingot was turned over and remelted several times to homogenize the samples. Energy-dispersive X-ray spectroscopy (EDX, SwiftED, Oxford) equipped with a scanning electron microscope  (TM3030, Hitachi High-Tech) was used to determine the composition of the samples (Co,Rh,Ir)Zr$_2$, (Fe,Co,Rh,Ir)Zr$_2$ and (Co,Ni,Cu,Rh,Ir)Zr$_2$. The configurational entropy of mixing ($\Delta S_{\rm{mix}}$) at the $Tr$ site was calculated from the composition of the $Tr$ site using the formula $\Delta S_{\rm{mix}}=-R\Sigma c_i \ln c_i$, where $R$ is the gas constant and $c_i$ is the mole fraction of component $i$. The crystal structure in high temperature was investigated by employing a powder XRD technique using the Cu-K$\rm{\alpha}$ radiation by the $\theta$-$2\theta$ method (Miniflex-600 RIGAKU) equipped with a heating temperature attachment (BTS 500). In this experiment, XRD patterns were analyzed using the Split pseudo-Voigt function of Toraya\cite{Toraya_1990}. The Rietveld refinement of the XRD data was performed using the RIETAN-FP package\cite{izumi_2007}. A schematic image was obtained by VESTA\cite{Momma_2011}. \par
The approximate composition of the $Tr$ site estimated by EDX for (Co,Rh,Ir)Zr$_2$, (Fe,Co,Rh,Ir)Zr$_2$ and (Co,Ni,Cu,Rh,Ir)Zr$_2$ were Co$_{0.28}$Rh$_{0.43}$Ir$_{0.29}$Zr$_2$, Fe$_{0.25}$Co$_{0.26}$Rh$_{0.28}$Ir$_{0.21}$Zr$_2$ and Co$_{0.19}$Ni$_{0.10}$Cu$_{0.08}$Rh$_{0.33}$Ir$_{0.31}$Zr$_2$, respectively. Using these calculations, the composition of Zr was fixed at 2. In this study, we abbreviated Co$_{0.28}$Rh$_{0.43}$Ir$_{0.29}$Zr$_2$ as (Co,Rh,Ir)Zr$_2$, Fe$_{0.25}$Co$_{0.26}$Rh$_{0.28}$Ir$_{0.21}$Zr$_2$ as (Fe,Co,Rh,Ir)Zr$_2$ and Co$_{0.19}$Ni$_{0.10}$Cu$_{0.08}$Rh$_{0.32}$Ir$_{0.31}$Zr$_2$ as (Co,Ni,Cu,Rh,Ir)Zr$_2$. Entropies of mixing of (Co,Rh,Ir)Zr$_2$, (Fe,Co,Rh,Ir)Zr$_2$ and (Co,Ni,Cu,Rh,Ir)Zr$_2$ were 1.08$R$, 1.42$R$ and 1.50$R$, respectively.\par
Figure 1(b) illustrates the XRD pattern and Rietveld refinement results for NiZr$_2$ at a temperature of 293 K. The obtained pattern was refined using the CuAl$2$-type (space group: \#140) structural model with an impurity phase of NiZr (9\%) concentration. The reliability factor was determined to be $R_{\text{wp}}$ = 4.1 \%, and the goodness-of-fit indicator was calculated as $S$ = 1.3. Figure 1(c) depicts the XRD at high temperatures, wherein the profiles at each temperature are similar, thus confirming the absence of structural transition. Figures 1(d) and (e) demonstrate that the 200 peak, indicative of the lattice constant $a$, and the 002 peak, indicative of the lattice constant $c$, respectively, shift to the lower angle side upon warming, indicating an expansion of the lattice constants $a$ and $c$. These results indicate the absence of the uniaxial NTE in NiZr$_2$. The Rietveld method was employed to obtain the lattice constants $a$ and $c$ and volume $V$ from the XRD results at each temperature, and their temperature dependence is summarized in Figures 1 (f), (g) and (h). Here, we estimate the linear thermal expansion coefficient (TEC), denoted as $\alpha$, using the formula
\begin{eqnarray}
\label{TEC}
\alpha= \left[\dfrac{1}{a(300 \,\textrm{K})}\right]\left(\dfrac{\Delta a}{\Delta T}\right).
\end{eqnarray}
 In this study, the entire temperature range was approximated as linear and fitted to the results. In CoZr$_2$, linear TEC along $a$- and $c$-axes are $\alpha_a=13$ $\rm{\mu K}^{-1}$ and $\alpha_c=11$ $\rm{\mu K}^{-1}$, respectively. These values are consistent with the results reported by Watanabe \textit{et al} ($\alpha_a=15.8$ $\rm{\mu K}^{-1}$, $\alpha_c=17$ $\rm{\mu K}^{-1}$).\cite{watanabe_2022}\par
Figure 2 depicts the high-temperature XRD patterns and the temperature dependence of lattice parameters for (Co,Rh,Ir)Zr$_2$, (Fe,Co,Rh,Ir)Zr$_2$ and (Co,Ni,Cu,Rh,Ir)Zr$_2$. In both (Co,Rh,Ir)Zr$_2$ and (Fe,Co,Rh,Ir)Zr$_2$, $Tr$Zr$_2$ (space group: \#227) impurity peaks were observed, in addition to $Tr$Zr$_2$ (space group: \#140), as illustrated in Figs. 2(a) and (e). Only the Rietveld analysis results at room temperature, with corresponding $R_{\textrm{wp}}$ values of 3.37 \% and 4.62 \%, were shown, as displayed in Figures 2(a) and 2(e). The Rietveld analysis revealed that the impurity content in (Co,Rh,Ir)Zr$_2$ was 7\% and 2\% in (Fe,Co,Rh,Ir)Zr$_2$. Conversely, no peaks indicative of impurities were detected in (Co,Ni,Cu,Rh,Ir)Zr$2$, as shown in Fig. 2(i), and the $R_{\textrm{wp}}$ in the Rietveld analysis at room temperature was 5.24 \%. As depicted in Figures 2(a), (e) and (i), the high-temperature XRD patterns of these materials remained unchanged, indicating the absence of structural changes. The peak of the 002 reflection, which indicates the lattice constant $c$, varied among the different materials, as shown in Figures 2(b), (f) and (j). As shown in Figures 2(b) and (j), the 002 peaks of (Co,Rh,Ir)Zr$_2$ and (Co,Ni,Cu,Rh,Ir)Zr$_2$ exhibited a slight shift towards higher angles as temperature increased, indicating a decrease in the lattice constant $c$ upon warming. Notably, the shift of the 002 peak towards the higher angle side with increasing temperature was particularly evident in (Fe,Co,Rh,Ir)Zr$_2$, as depicted in Figure 2(f). As shown Figure 2 (c), the $c$-axis of (Co,Rh,Ir)Zr$_2$ exhibits uniaxial NTE over the entire temperature range measured. Its lattice constant $a$ slightly decreases at room temperature and exhibits PTE above 370 K. The minor decrease in the a-axis near room temperature is hypothesized to be due to amorphization, but the underlying cause remains unclear. The temperature dependence of the volume was NTE near room temperature and PTE above 400 K. Given that volumetric NTE has been reported near room temperature for CoZr$_2$, it is likely that the room temperature volumetric NTE of (Co,Rh,Ir)Zr$_2$ is a result of the presence of CoZr$_2$ as depicted by the green closed circle in Figure 3(c)\cite{Mizugu_2022_1}. The lattice constants $a$ and $c$ of the (Fe,Co,Rh,Ir)Zr$_2$ exhibit expansion and contraction, respectively, and the temperature dependence of volume $V$ is increased. The temperature dependence of the lattice constants and volume of the (Co,Ni,Cu,Rh,Ir)Zr$_2$ is similar to those of (Fe,Co,Rh,Ir)Zr$_2$, however, the change ration in the lattice constants $c$ is less pronounced in comparison to (Fe,Co,Rh,Ir)Zr$_2$. \par
In order to compare the magnitude of change in the lattice constants and volume, Figure 3 illustrates the temperature dependence of the change ratio of lattice constants in the $a$-axis ($\Delta a/ a= 1- a(300 {\rm K})/a(T)$), $c$-axis ($\Delta c/ c= 1-c(300 {\rm K})/c(T)$) and volume ($\Delta V/ V= 1-V(300 {\rm K})/V(T)$) normalized at room temperature of $Tr$Zr$_2$ ($Tr$ = Ni, Co, (Fe,Co,Ni), (Co,Rh,Ir), (Fe,Co,Rh,Ir), (Fe,Co,Ni,Rh,Ir), (Co,Ni,Cu,Rh,Ir), (Fe,Co,Ni,Cu,Ga))\cite{Mizugu_2022_1,Kasem_2022}. Table 1 summarizes the linear TEC of these materials,  calculated using equation (1). As depicted in Figure 3(a), the lattice constants $a$ increase for all materials, and the change ratio is comparable across all of them. As shown in Table 1, the linear TEC along the $a$-axis is of the order of +10 $\mu\rm{K}^{-1}$. Exceptions include $Tr=$(Co, Rh, Ir), which $a$-axis exhibit contraction near room temperature and a relatively large linear TEC $\alpha_a$=+30 $\mu\rm{K}^{-1}$ above 450 K. In contrast, the change ratio of the $c$-axis varies among materials as depicted in Figure 3(b). The $c$-axis of (Fe,Co,Ni,Cu,Ga)Zr$_2$ exhibited almost temperature independence, while that of NiZr$_2$ displayed expansion; other compounds shows $c$-axis contraction. As show in Table 1, the linear TEC along the $c$-axis is smallest for CoZr$_2$ ($\alpha_c$=-28 $\mu\rm{K}^{-1}$ above 300 K) and increases in the order of (Fe,Co,Rh,Ir), (Fe,Co,Ni), (Fe,Co,Ni,Rh,Ir), (Co,Rh,Ir), (Co,Ni,Cu,Rh,Ir).  As depicted in Figure 3(c), the change ratio in volume $\Delta V/V$ increases with increasing temperature, with the exception of $Tr$=Co, (Co,Rh,Ir) as shown in Figure 3(c). The material with the greatest volumetric change is NiZr$_2$, while the material with the least volumetric change is (Fe,Co,Ni,Rh,Ir)Zr$_2$ . The reason for the lower volumetric change in (Fe,Co,Ni,Rh,Ir)Zr$_2$ is thought to be due to the fact that the absolute value of the linear TEC along the $c$-axis $\alpha_c$ is significantly greater than that of the $a$-axis $\alpha_a$ by 45\%, resulting in offsetting expansion in the $a$-axis by contract along the $c$-axis. The absolute value of $\alpha_c$ of (Fe,Co,Rh,Ir)Zr$_2$ is also larger than $\alpha_a$, but only by about 0.4 \%, thus the effect of $c$-axis contract on the volume change is smaller and the volume expansion rate is likely greater. The temperature dependence of the volume of CoZr$_2$ and (Co,Rh,Ir)Zr$_2$ exhibits a similar trend, with the volume decreasing near room temperature and increasing at higher temperatures. \par

To discuss the effect of uniaxial NTE with the $c$-axis contraction in $Tr$Zr$_2$ compounds, we demonstrated the temperature dependence of the ratio of the $c$- and $a$-axis lattice constants ($c/a$) for $Tr$Zr$_2$ ($Tr$ = Ni, Co, (Fe,Co,Ni),  (Co,Rh,Ir), (Fe,Co,Rh,Ir), (Fe,CoNi,Rh,Ir), (Co,Ni,Cu,Rh,Ir), and (Fe,Co,Ni,Cu,Ga)) in Figure 4 (a). In the uniaxial NTE compounds ($Tr=$Co, (Fe,Co,Ni),  (Co,Rh,Ir), (Fe,Co,Rh,Ir), (Fe,CoNi,Rh,Ir), and (Co,Ni,Cu,Rh,Ir)), $c/a$ decreases with increasing temperature due to the contraction and expansion of the lattice constants $c$ and $a$, respectively. NiZ$_2$ and (Fe,Co,Ni,Cu,Ga)Zr$_2$ showed a smaller change in $c/a$ than other compounds, owing to the expansion of the lattice constant $c$. In addition, room temperature $c/a$ values were found to be larger for uniaxial NTE compounds than for PTE compounds. \par

In Figure 4(b), we depict the linear TEC along the c-axis $\alpha_c$ as a function of the $c/a$ ratio at room temperature. Figure 4(b) clearly illustrates the relationship between $\alpha_c$ and $c/a$, with $\alpha_c$ increasing as $c/a$ decreases, and the sign of $\alpha_c$ switching from negative to positive at a threshold of approximately $c/a \sim$ 0.83 by manipulating $c/a$. During this experiment, the mixing entropy $\Delta S_{\rm{mix}}$ at the $Tr$ sites varied, which may have influenced the temperature dependence of the lattice constant $c$. This experiment does not rule out the possibility that the disparity in $\Delta S_{\rm{mix}}$ at the $Tr$ sites affected the electronic state and resulted in anisotropic thermal expansion. However, we claim that the association between $\Delta S_{\rm{mix}}$ and anisotropic thermal expansion is insignificant, as CoZr$_2$ and NiZr$_2$ with $\Delta S_{\rm{mix}}$=0 exhibit contrasting thermal expansions, while (Fe,Co,Ni,Rh,Ir)Zr$_2$ ($\Delta S_{\rm{mix}}$=1.60$R$) and (Co,Ni,Cu,Rh,Ir)Zr$_2$ ($\Delta S_{\rm{mix}}$=1.50$R$) with relatively high $\Delta S_{\rm{mix}}$ display dissimilar temperature dependencies of the lattice constant $c$. Notably, the anisotropy of thermal expansion is more dependent on the $c/a$ ratio than the type of $Tr$-site atoms, implying that the $c$-axis thermal expansion can be regulated by flexible substitution of the $Tr$ sites in $Tr$Zr$_2$.\par 
Materials exhibiting anisotropic thermal expansion over a wide temperature range, with expansion in one axial direction and contraction in another, include hexagonal boron nitride (hBN) and BiB$_3$O$_6$\cite{Paszkowicz_2002, Tenga_2002}. The linear TECs along the $a$- and $c$-axes of hBN near room temperature are $\alpha_a=-2.72$ $\mu$K$^{-1}$ and $\alpha_c=+37.3$ $\mu$K$^{-1}$, respectively. Anisotropic bonding energy have been proposed as the origin of this phenomenon. Conversely, BiB$_3$O$_6$ has TECs of $\alpha_a=-26.4$ $\mu$K$^{-1}$, $\alpha_b=+50.4$ $\mu$K$^{-1}$, and $\alpha_c=+8.5$ $\mu$K$^{-1}$, and the theoretical origin of its anisotropic thermal expansion has been proposed to be chiral acoustic phonons with elliptical motion of the bismuth atoms, based on first-principles calculations\cite{Romao_2019}. In the future, it is important to investigate the temperature dependencies of the lattice parameters of various $Tr$Zr$_2$, including HEAs, to unveil the systematic relationship between crystal structure and thermal expansion. To clarify the origin of the anisotropy of thermal expansion of the $c$-axis, it is important to elucidate the local crystal structure by high-resolution XRD and NPD.

In conclusion, we synthesized transition-metal zirconides NiZr$_2$, (Co,Rh,Ir)Zr$_2$, (Fe,Co,Rh,Ir)Zr$_2$, and (Co,Ni,Cu,Rh,Ir)Zr$_2$ with a CuAl$_2$-type structure (space group \#140) and analyzed their crystal structures at high temperatures using powder XRD. We observed that while all compounds exhibited PTE in the lattice constant $a$, the temperature dependence of the $c$-axis varied among the compounds. In particular, NiZr$_2$ demonstrated PTE, while NTE was observed for (Co,Rh,Ir)Zr$_2$, (Fe,Co,Rh,Ir)Zr$_2$, and (Co,Ni,Cu,Rh,Ir)Zr$_2$. Our findings indicate that the temperature dependence of the lattice constant $c$ is correlated with the ratio of the $c$-axis to the $a$-axis ($c/a$), and that manipulating $c/a$ through elemental substitution at the $Tr$ site can switch between NTE and PTE. Although the underlying mechanism for this anisotropic lattice structural change remains unclear, this study provides a promising avenue for the development of ZTE compounds.\par
\section{acknowledgment}
The authors would like to thank Y. Watanabe, M. Fujita and O. Miura for their support in experiments and discussion. This work was partly supported by Grant-in-Aid for Scientific Research (KAKENHI) (Proposal No. 21K18834 and 21H00151) and Tokyo Government Advanced Research (H31-1).
\begin{figure}[H]
\begin{center}
\includegraphics[scale=0.95]{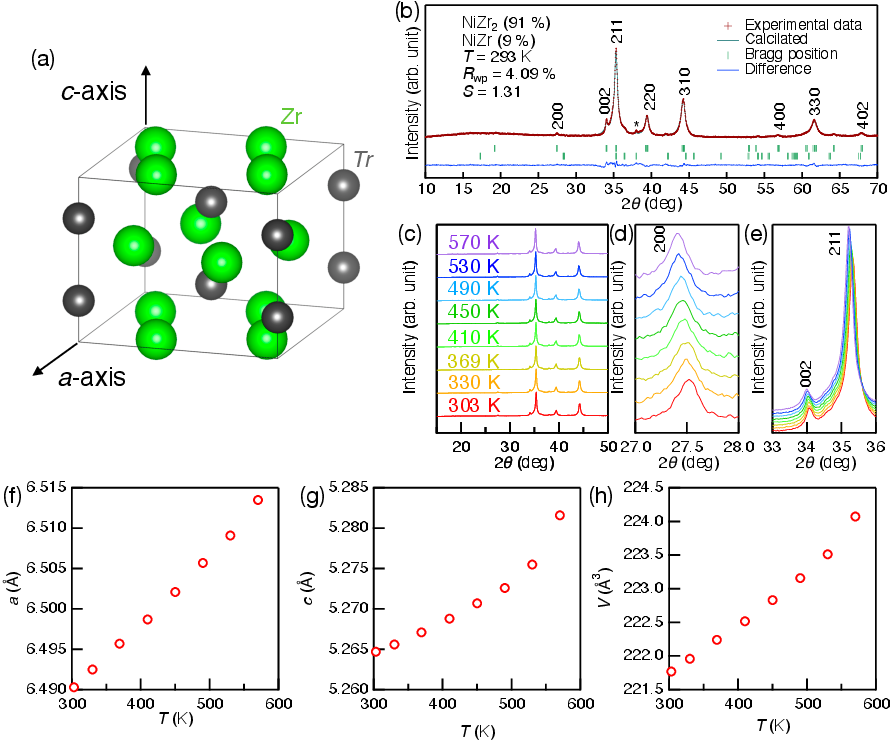}
\caption{(Color online) (a) A schematic image of the tetragonal crystal structure of $Tr$Zr$_2$. (b)Observed ($+$ symbols), calculated (upper green solid line), and difference (lower solid line) patterns of NiZr$_2$ at $T$=293 K obtained by Rietveld refinement from XRD pattern data. The green ticks upwards and downwards represent the positions of possible Bragg reflection of NiZr$_2$ and impurity of NiZr, respectively. The numbers indicate to the Miller indices of NiZr$_2$. An asterisk denotes an impurity peak. (c) XRD patterns at $T$= 303 , 330 , 369 , 410 , 450 , 490 , 530, and 570 K. (d)(e)XRD patterns near the 200 and 002, respectively, with colors corresponding to temperatures in (c). (f-h) Temperature dependence of lattice constant $a$, $c$ and $V$ of NiZr$_2$.}
\label{struct}
\end{center}
\end{figure}
\begin{figure}[H]
\begin{center}
\includegraphics[scale=0.6]{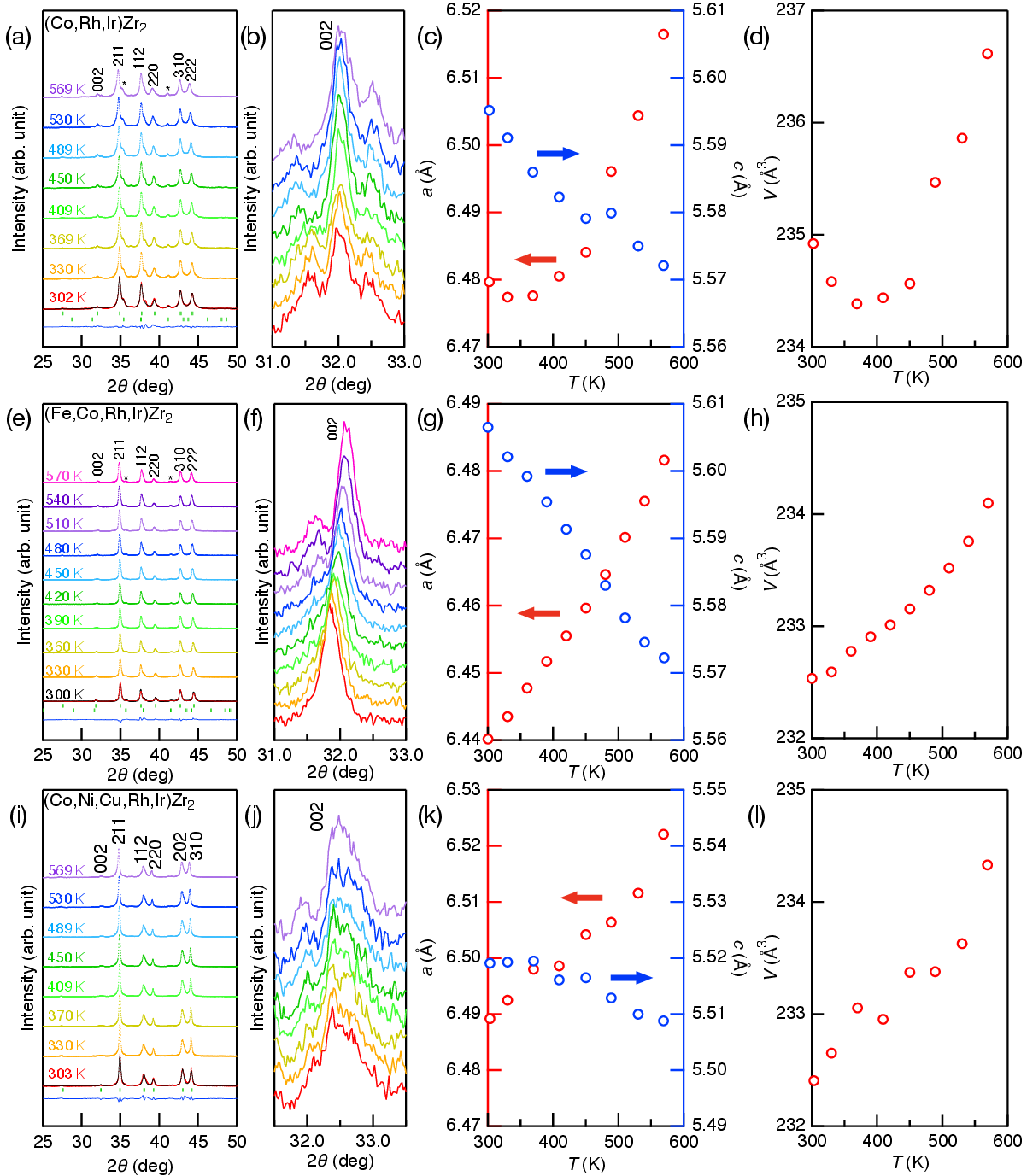}
\caption{(Color online) (a)(e)(i) X-ray Diffraction (XRD) patterns at elevated temperatures; (b)(f)(j) XRD patterns in the vicinity of the 002 reflection, with colors corresponding to the temperatures in (a),(e), and (i), respectively; (c)(g)(k) temperature-dependent lattice constants of the a and c axes; (d)(h)(l) temperature-dependent volume V for (Co,Rh,Ir)Zr2 (upper panels), (Fe,Co,Rh,Ir)Zr2 (middle panels), and (Co,Ni,Cu,Rh,Ir)Zr2 (lower panels), respectively. (a)(e)(i) The results of Rietveld analysis are only presented for room temperature (indicated by red markers). The calculated value is represented by a black solid line, and the difference is shown by a blue solid line. The Bragg peaks are indicated by green ticks, and the numbers correspond to the Miller indices of NiZr2. An asterisk indicates an impurity peak. In (a)(e), the upward green ticks represent the Bragg peaks of the sample, while the downward green ticks represent the Bragg peaks of the impurity.}
\label{NiZr2_XRD}
\end{center}
\end{figure}

\begin{figure}[H]
\begin{center}
\includegraphics[scale=1]{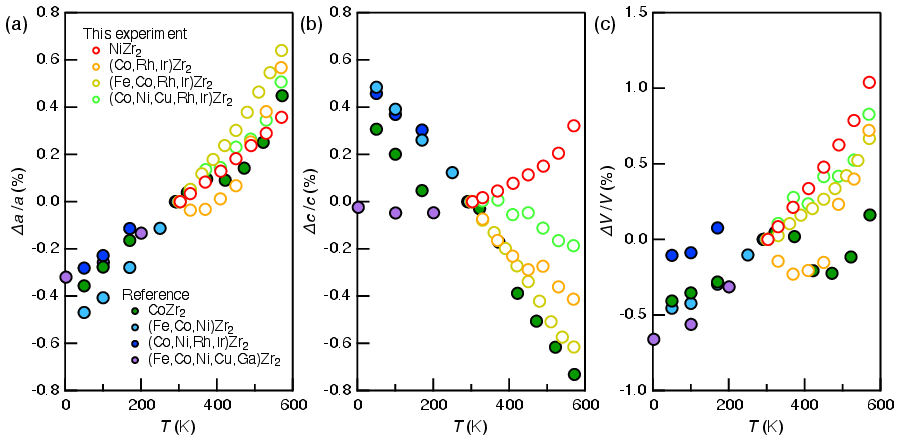}
\caption{(Color online) Temperature dependence of the change ratio of (a) $a$-axis and (b) $c$-axis lengths and (c) volume $V$ scaled at room temperature in $Tr$Zr$_2$ ($Tr$=Ni, Co, (Fe,Co,Ni),  (Co,Rh,Ir), (Fe,Co,Rh,Ir)Zr$_2$, (Fe,CoNi,Rh,Ir), (Co,Ni,Cu,Rh,Ir), (Fe,Co,Ni,Cu,Ga)). The results obtained in this experiment are represented by open circles, while closed circles are sourced from references (Ref. 17 ($Tr$=Co, (Fe,Co,Ni), and (Co,Ni,Cu,Rh,Ir)) and Ref.18 ($Tr$=(Fe,Co,Ni,Cu,Ga)).}
\label{rel_change}
\end{center}
\end{figure}

\begin{table}[H]
\begin{center}
\caption{Linear and volumetric TEC for $Tr=$Ni, Co, (Fe,Co,Ni),  (Co,Rh,Ir), (Fe,Co,Rh,Ir). (Fe,CoNi,Rh,Ir), (Co,Ni,Cu,Rh,Ir), (Fe,Co,Ni,Cu,Ga). Their values for $Tr=$Co, (Fe,Co,Ni), (Fe,Co,Ni,Rh,Ir), and (Fe,Co,Ni,Cu,Ga) refer to Ref. 17 and Ref. 18.}
\begin{tabular}{cccc}
\hline
$Tr$             & $\alpha_a$ ($\rm{\mu K}^{-1}$) & $\alpha_c$ ($\rm{\mu K}^{-1}$) & Reference            \\ \hline
Ni               & +13                            & +11                            & this study           \\ \hline
                 & +18 (0 K$< T <$ 300 K)          & -15 (0 K$<T<$300 K)          &        Ref. 17              \\
Co               & +14 (300 K$<T$)               & -28 (300 K$<T$)               &  \\ \hline
(Co,Rh,Ir)       & -12 (300 K$<T<$ 370 )        & -14                            & this study           \\
                 & +30 (450 K$<T$)               &                                &                      \\ \hline
(Fe,Co,Ni)       & +19                            & -19                            &Ref. 17 \\ \hline
(Fe,Co,Rh,Ir)    & +24                            & -25                            & this study           \\ \hline
(Co,Ni,Cu,Rh,Ir) & +17                            & +7                             & this study           \\ \hline
(Fe,Co,Ni,Rh,Ir) & +11                            & -16                            & Ref. 17 \\ \hline
(Fe,Co,Ni,Cu,Ga) & +11                            & +0.7                           &Ref. 18   \\ \hline
\end{tabular}
\end{center}
\end{table}

\begin{figure}[H]
\begin{center}
\includegraphics[scale=0.7]{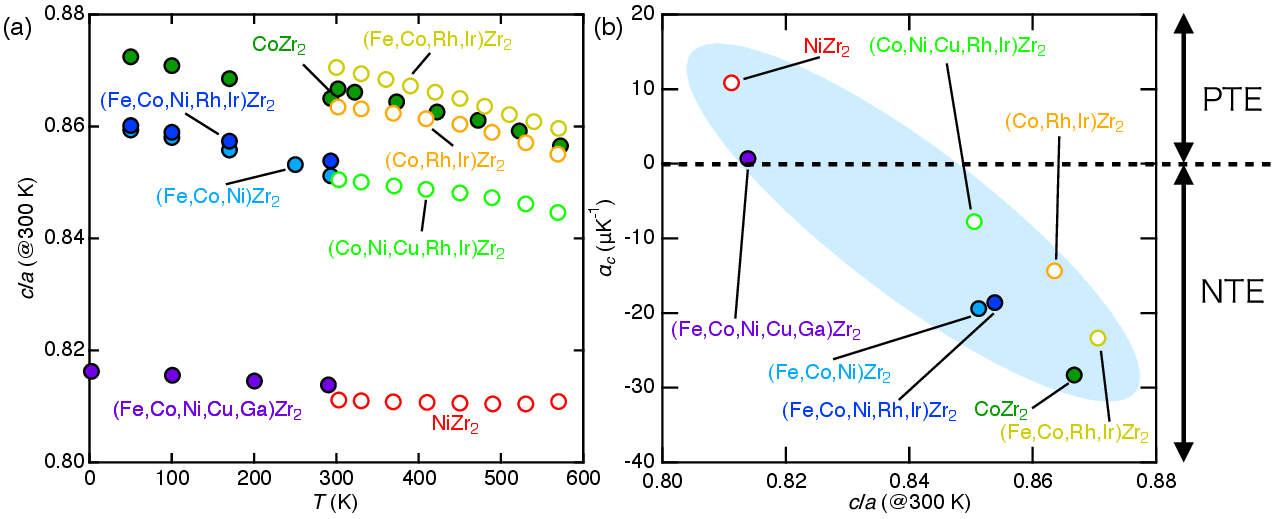}
\caption{(Color online) (a) Temperature dependence of the ratio of lattice constants $c$ and $a$ ($c/a$) and (b) the linear TEC along the $c$-axis $\alpha_c$ plotted as a function of $c$/$a$ for $Tr$Zr$_2$ ($Tr$=Ni, Co, (Fe,Co,Ni),  (Co,Rh,Ir), (Fe,Co,Rh,Ir). (Fe,CoNi,Rh,Ir), (Co,Ni,Cu,Rh,Ir), (Fe,Co,Ni,Cu,Ga)). The value of $c$/$a$ is taken from Ref.17 ($Tr$=Co, (Fe,Co,Ni), (Co,Ni,Cu,Rh,Ir)) and Ref.18($Tr$=(Fe,Co,Ni,Cu,Ga)). }
\label{c_a}
\end{center}
\end{figure}

\end{document}